\documentclass[prl,twocolumn,superscriptaddress,showpacs,preprintnumbers,amsmath,amssymb]{revtex4}
\usepackage{graphicx,float}
\usepackage{epstopdf}
\usepackage{dcolumn}
\usepackage{bm}
\usepackage{color}

\let\ol=\overline
\let\dd=\partial

\let\f=\frac

\newcommand{\be}{\begin{equation}}
\newcommand{\ee}{\end{equation}}
\newcommand{\bc}{\begin{center}}
\newcommand{\ec}{\end{center}}
\newcommand{\bq}{\begin{quote}}
\newcommand{\eq}{\end{quote}}

\newcommand{\bu}{\bf{u}}
\newcommand{\bg}{\bf g}

\newcommand{\dr}{\left(\frac{\Delta\rho}{\rho_s}\right)}
\newcommand{\nab}{\mbox{\boldmath $\nabla$} {}}
\newcommand{\ui}{\overline u_i}
\newcommand{\uj}{\overline u_j}

\newcommand{\odr}{\left(\frac{\Delta\overline{\rho}}{\rho_s}\right)}

\begin{document}

\newcommand{\LANL}{Condensed Matter \& Thermal Physics Group and Center for Nonlinear Studies,\\Los Alamos National Laboratory, Los Alamos, New Mexico 87545, USA}
\newcommand{\Lyon}{Laboratoire de Physique,
\'Ecole Normale Sup\'erieure de Lyon,
46, all\'ee d'Italie 69364 Lyon Cedex 07, FRANCE}

\title{Mixing in stratified gravity currents: Prandtl mixing length}

\author{P. Odier}
\affiliation{\LANL}
\affiliation{\Lyon}

\author{J. Chen}
\affiliation{\LANL}

\author{M. K. Rivera}
\affiliation{\LANL}

\author{R. E. Ecke}
\affiliation{\LANL}

\date{\today}

\begin{abstract}
Shear-induced vertical mixing in a stratified flow is a key ingredient of thermohaline circulation.  We experimentally determine the vertical flux of 
momentum and density of a forced gravity current using high-resolution velocity and density measurements.  
A constant eddy viscosity model provides a poor description of the physics of mixing, but a Prandtl mixing
length model relating momentum and density fluxes to mean velocity and density gradients works well. For $\langle Ri_g \rangle \approx 0.08$ and $Re_\lambda \approx 100$, the mixing lengths are fairly constant, about the same magnitude, comparable to the turbulent shear length.
\end{abstract}

\pacs{47.20.Ft,47.27.Wj,47.55.Hd,92.10.Lg}
\maketitle

Mixing in stratified shear flows is an important process in many geophysical situations including
atmospheric shear layers or the upper ocean mixing induced
by wind stresses at the surface \cite{DeSilva96,Riley00,Thorpe07}.  Of particular current interest are the mixing and entrainment of oceanic overflows, which are involved in the transport of heat and salt in the global ocean via the thermohaline ``conveyor belt" \cite{Willebrand01, Hansen00, BAcon98, Price93}. Such circulations are thought to play a significant role in decadal  predictability of ocean evolution.  Understanding the physics of mixing in stratified layers and providing a simple description of this mixing may help improve predictions of global climate change \cite{Winton98}.

Laboratory studies of stratified mixing layers \cite{Thorpe73,Koop79,Strang01} and gravity currents on shallow inclines \cite{Ellison59,Baines05,Decamp07} have characterized the mixing and entrainment resulting from the competition of shear and buoyancy.  Whereas fundamental studies of shear layers have provided reasonable characterization of some turbulence quantities, the gravity current experiments have focused on bulk entrainment measurements rather than on details of the turbulence itself.  There is a need to apply modern turbulence measurements to the physically relevant problem of inclined gravity currents as a model of turbulent oceanic overflows. In particular, understanding turbulent mixing implies being able to describe how correlations of small scale fluctuating quantities affect large-scale fluid transport properties.

We developed an experiment of a stratified flow on an inclined plane that is destabilized by shear. Our main result is that the turbulent transport of momentum and density are described in a direct and compact form by a Prandtl mixing length
model \cite{Prandtl,Hinze75}.  In particular, the turbulent vertical fluxes of momentum and density scale with the vertical mean gradients of velocity $\partial_z \ol u$ and density $\partial_z \ol\rho$ as an eddy viscosity $\nu_T = L_m^2 \vert\partial_z \ol u\vert$ and an eddy diffusivity $\gamma_T = L_\rho^2 \vert\partial_z \ol u\vert$ where $L_m$ and $L_\rho$ are approximately constant over the mixing zone of the stratified shear layer. In general, Prandtl mixing length models, although widely used because of their simplicity, are often not verified by simulation or experiment \cite{Pope}. For example, a mixing
length description does not work well in our system when the flow is unstratified, suggesting that stratification plays a key role in the correlations we measure.  More complicated descriptions beyond a simple eddy-viscosity or mixing length approach have been developed \cite{Pope} but seem unnecessary here.  Our compact description may provide an efficient parameterization of mixing and entrainment in oceanic overflows.

\begin{figure}[h]
{\begin{minipage}{.47\textwidth}
\centering
\includegraphics[width=8.5cm]{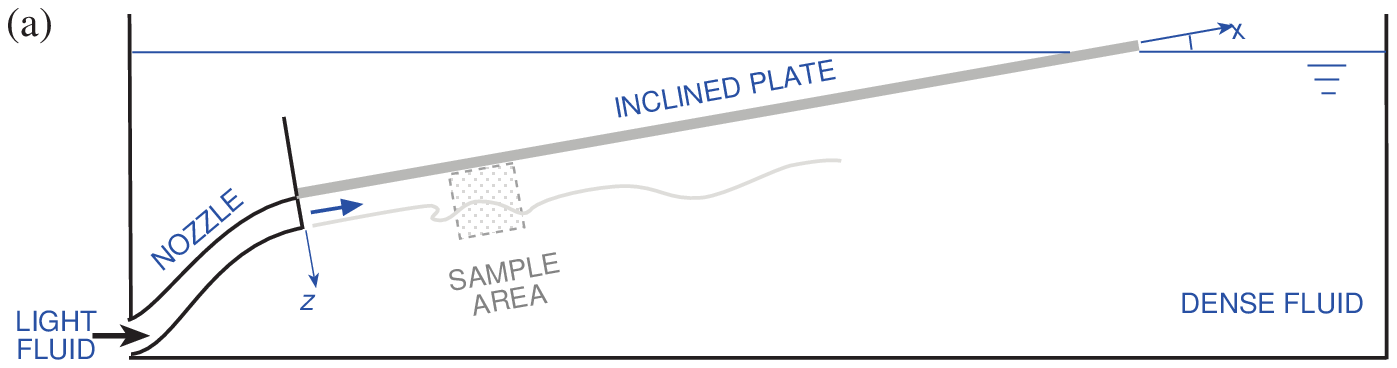}\\
\vspace{0.1cm}
\includegraphics[width=9cm]{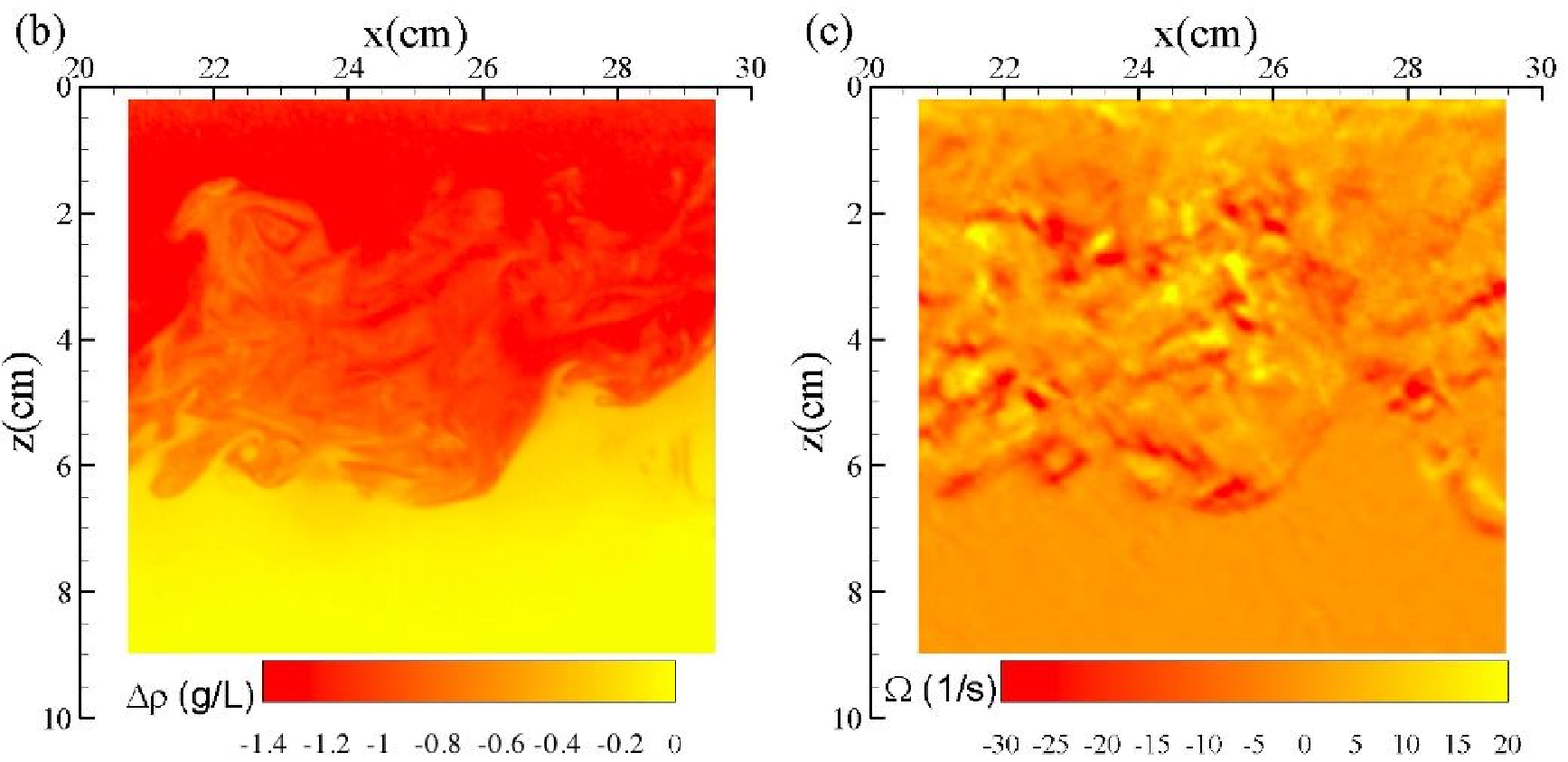}\\
\caption{\label{fig:images} (Color Online) a) Sketch of the experimental device. b) Density snapshot obtained from PLIF and c) vorticity snapshot from 
PIV. The scale of density is from 0 (yellow, salt water) to -1.4 g/L (red, mixture of salt water with ethanol). The scale of vorticity is from positive (yellow) to negative (red) with vorticity amplitude in the range $-12 < \Omega_y < 6$ s$^{-1}$.
}
\end{minipage}}
\end{figure}

The experiment, sketched in Fig.~\ref{fig:images}a, and described in detail elsewhere \cite{asme_proc}, consists of a turbulent, uniform-density flow injected via a pump through a 5 cm high by 45 cm wide nozzle at a speed of $U_0$ = 8 cm/s into a tank filled with unstirred higher density fluid.  The turbulence level of the injection current is enhanced by an active grid device located just before the injection nozzle. The flow, upon exiting the nozzle, is bounded from above by a transparent plate inclined at an angle of 10$^{\rm o}$ with respect to horizontal, is unbounded below, and is confined in a tank about 2 m long, 0.5 m wide and 0.5 m high.  The components of the spatial position vector ${\bf x}$ describing the flow are the mean flow direction $x$, the cross-stream direction $y$ and the downward distance perpendicular to the plate $z$. The corresponding velocity ${\bf u(x)}$ has components $\lbrace u,v,w\rbrace$. We use the notation ${\bf \ol u}$ for a time- and ensemble-averaged quantity and ${\bf u'=u-\ol u}$ for its fluctuating part. The exit fluid, a solution of ethanol and water, is less dense ($\rho_e = 996.8$ g/L) than the fluid in the tank, water and salt (NaCl) ($\rho_s =999.4$ g/L).  The net density difference between the fluids is $\Delta \rho = 2.6$ g/L. The concentrations of ethanol and salt are adjusted (and the fluid temperatures maintained equal within 0.2$^\circ$C) so that the fluids are index matched to avoid optical distortions\cite{McDougall}. All the fluids are freshly prepared for each run, which lasts for about 45 s.

Instantaneous velocity and density fields are measured in a 9 cm $\times$ 9 cm area of a 0.1 cm thick laser sheet in the $x-z$ plane.  Velocity and density are measured simultaneously using particle image velocimetry (PIV) and planar laser-induced fluorescence (PLIF), respectively, at a rate of 3 Hz with two 2048$^2$ pixels digital cameras.  Fluorescent dye (Rhodamine 6G) is added to the light fluid, and a calibration of density versus fluorescence intensity is performed for each position of the field of view. Figure~\ref{fig:images} shows a snapshot of the density difference obtained from the PLIF 
and of the $y$-component of vorticity, derived from the PIV velocity field 
($\Omega_y=\partial_z u(x,z) - \partial_x w(x,z)$). 

The lighter exit fluid is stably stratified with respect to the heavy fluid in the tank and forms a gravity current on the bottom side of the plate. The competition between the stabilizing effect of buoyancy and the destabilizing shear is captured in a dimensionless parameter, the gradient Richardson number, $Ri = -(g/\rho_s)(\partial_z \ol\rho)/ (\partial_z \ol u)^2$ where $g$ is acceleration of gravity.  For small $Ri$, shear dominates buoyancy, and the flow is unstable to Kelvin-Helmoltz instability \cite{Thorpe07}.  The gravity current is fully turbulent as it exits the nozzle with streamwise velocity fluctuations $u'$ about 25\% of $\ol u$, corresponding to a Taylor Reynolds number $Re_\lambda = u'^2/\sqrt{15\epsilon\nu} \approx 100$, where $\nu$ is the fluid kinematic viscosity and $\epsilon$ is the mean dissipation rate measured directly from velocity field (the spatial resolution of our velocity measurement is 0.5 mm compared to the dissipation scale of 0.33 mm). 
Over the first 20 cm, there is rapid evolution of mean quantities including the vertical-velocity and density gradients. In this Letter, we focus on the region from 21 to 45 cm over which averages are approximately stationary, e.g.,  $Ri \approx 0.08$, 
$\partial_z \ol u \approx 1\ s^{-1}$, $(1/\ol\rho)\partial_z \ol\rho \approx 10^{-4}$ cm$^{-1}$, and $\epsilon \approx 1\ cm^2/s^3$.  Note, however, that the results decribed here also apply to the initial region, except for a stronger dependence on downstream distance. Details of the spatial distribution of mean and fluctuating quantities in vertical and streamwise directions will be presented elsewhere.

\begin{figure}[h]
{\begin{minipage}{.47\textwidth}
\includegraphics[width=4.1cm]{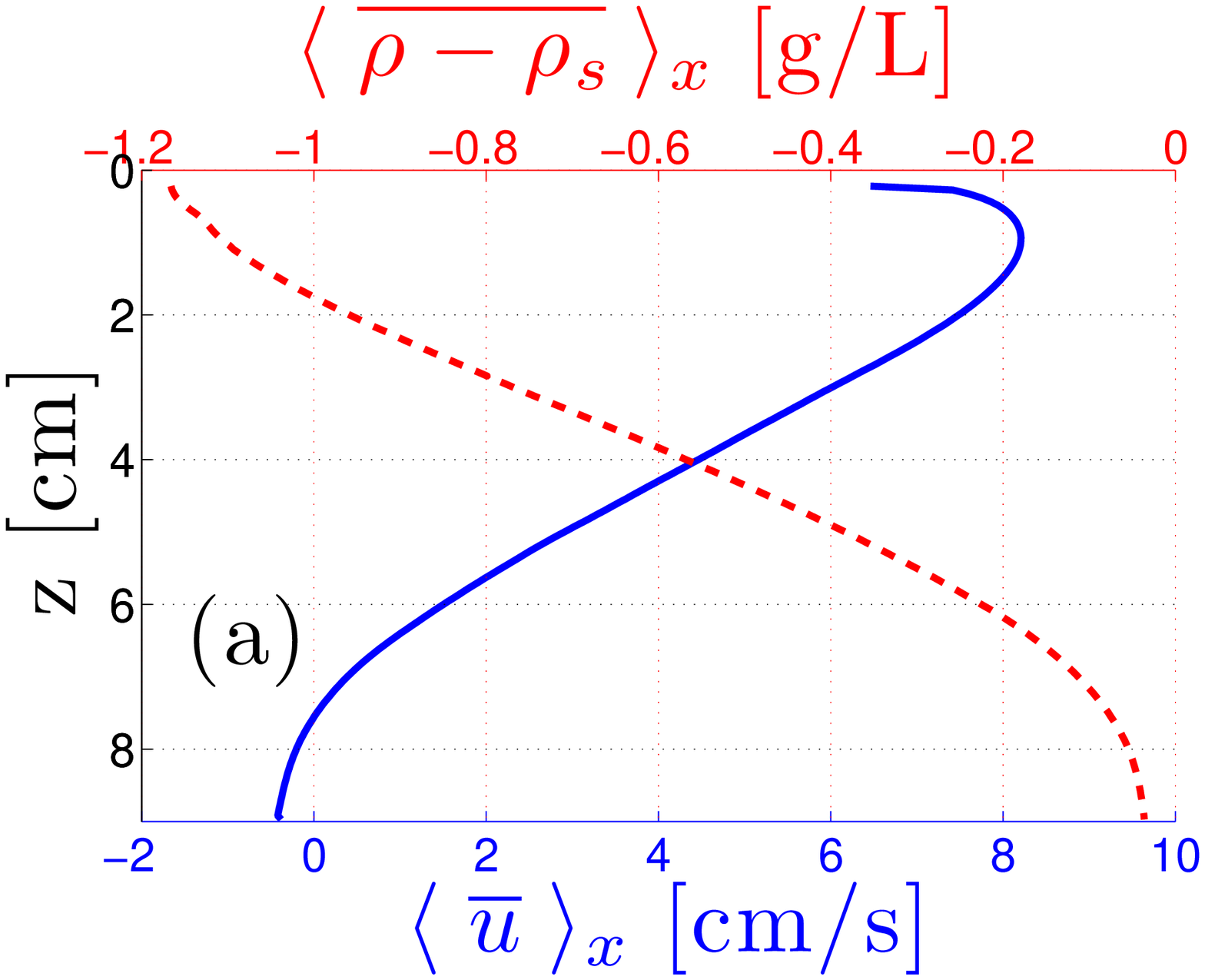}
\vspace{0.2 cm}
\includegraphics[width=4.1cm]{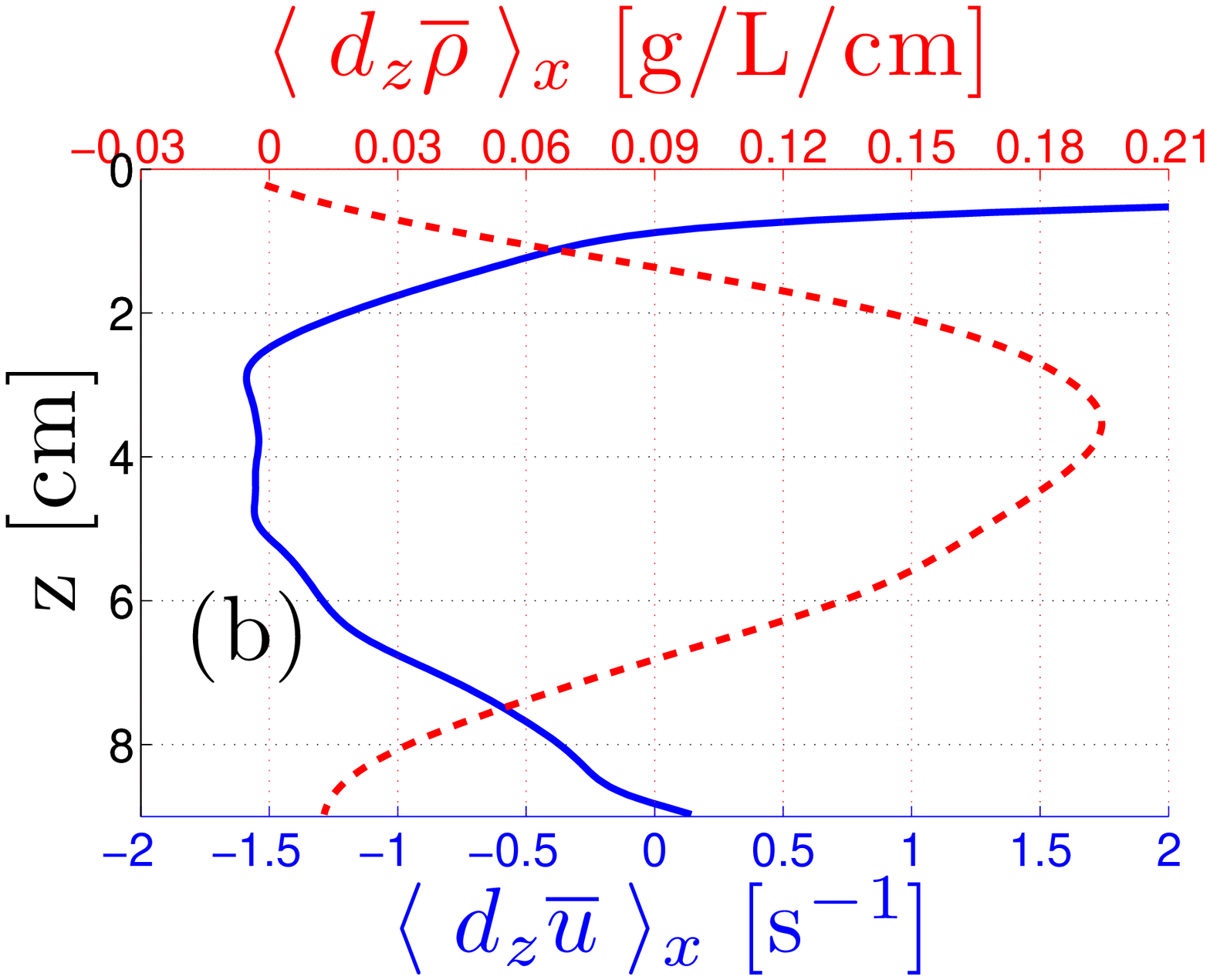}\\
\caption{\label{fig:mean} (Color Online) The time averaged values of a) down-plane velocity $\ol u$ (solid line, bottom axis) and density difference $\ol{\Delta \rho}$ (dashed, top axis) and b) the vertical gradients $d_z\ol u$ (solid, bottom axis) and $d_z\ol\rho$ (dashed, top axis) as functions of the distance $z$ from the plane. These profiles have also been averaged along the downstream distance ($\langle\bullet\rangle_x$ symbol), between 21 and 47 cm.}
\end{minipage}}
\end{figure}

The time-averaged profiles of density difference $\ol{\Delta\rho}=\ol{\rho-\rho_s}$ and downstream velocity $\ol{u}$ as functions of $z$ are shown in Fig.~\ref{fig:mean}a. Figure~\ref{fig:mean}b shows the corresponding profiles of $\partial_z\ol\rho$ and $\partial_z\ol u$. There is a strong mixing region between 1.5 cm $<  z  < $7 cm where the gradients are within 50\% of their maximum values. The velocity field goes to zero over a turbulent boundary layer (not resolved) which leads to the velocity maximum at $z \approx 1$ cm.  The vertical density gradient decreases near the wall indicating less vigorous mixing owing to the presence of the boundary. Nevertheless, some mixing has
occurred at lower $x$, as indicated by the reduced density difference at the wall compared to the initial difference. Far from the wall, the velocity and density difference approach quiescent values because of the stabilizing influence of stratification.

The density and momentum evolution equations are:
\begin{eqnarray*}\label{equ_syst}
 \label{equ_mom} \partial_t\bu+\left({\bu}\cdot{\nab}\right){\bu} & = & -\f{1}{\rho}{\nab} p+\nu\nab^2{\bu}-{\bg}\dr\\
\label{equ_dens}  \partial_t\dr+\left({\bu}\cdot{\nab}\right)\dr & = &\kappa\nab^2\dr\;\;,
\end{eqnarray*}

\noindent where $p$ is the pressure field and $\kappa$ is the molecular mass diffusivity of the light fluid into the heavy one. For a turbulent flow, the Reynolds decomposition of quantities into mean and fluctuating parts yields an additional effective force density due to the turbulent fluctuations \cite{Pope}. It takes the form of the divergence of the stress tensor $\rho\ol{u'_iu'_j}$. In the same way, for a flow with variable density, the divergence of  $\ol{\rho'\bf u'}$ appears in the density equation. The equations for the evolution of the averaged momentum and density, in the stationary case, become (with the spatial derivative along the i$^{\rm th}$ component of $\bf x$ denoted $\dd_i$):
\begin{eqnarray*}\label{equ_syst2}
 \label{equ_mom2} \noindent\uj\dd_j\ui & = & -\f{1}{\rho}\dd_i \overline p+\nu\dd_{jj}\ui-\dd_j\overline{u_i'u_j'}-g_i\odr\;\;\;\;\\
\label{equ_dens2}\uj\dd_j\odr & = &\kappa\dd_{jj}\odr-\f{1}{\rho_s}\dd_j\overline{\rho'u_j'}\;\; ,
\end{eqnarray*}

In particular,  $\ol{u'w'}$ and $\ol{\rho'w'}$ can be interpreted, respectively, as the vertical flux of downstream momentum and of density due to turbulent fluctuations. To understand the transport mechanisms that maintain the vertical gradients shown in Fig.\ \ref{fig:mean}, we need to relate the turbulent fluxes
to the mean gradients. A closure scheme commonly used in turbulence establishes these relations via effective diffusivity coefficients (or eddy diffusivities) $\nu_T=-\ol{u'w'}/\dd_z\ol u$ and $\gamma_T=-\ol{\rho'w'}/\dd_z\ol\rho$.

\begin{figure}[h]
{\begin{minipage}{.5\textwidth}
\centering
\includegraphics[width=5.4cm]{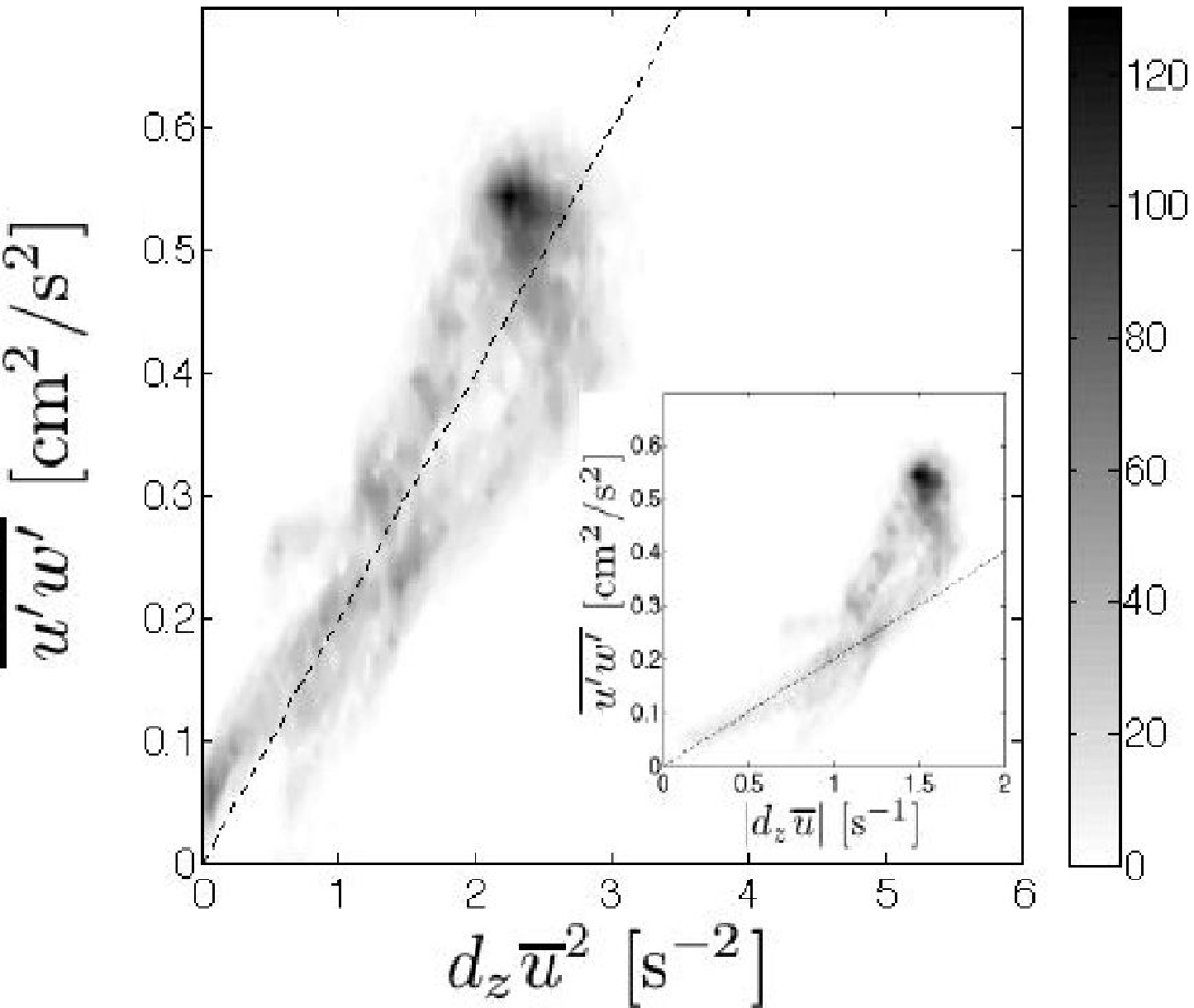}\\
\vspace{0.4 cm}
\includegraphics[width=5.4cm]{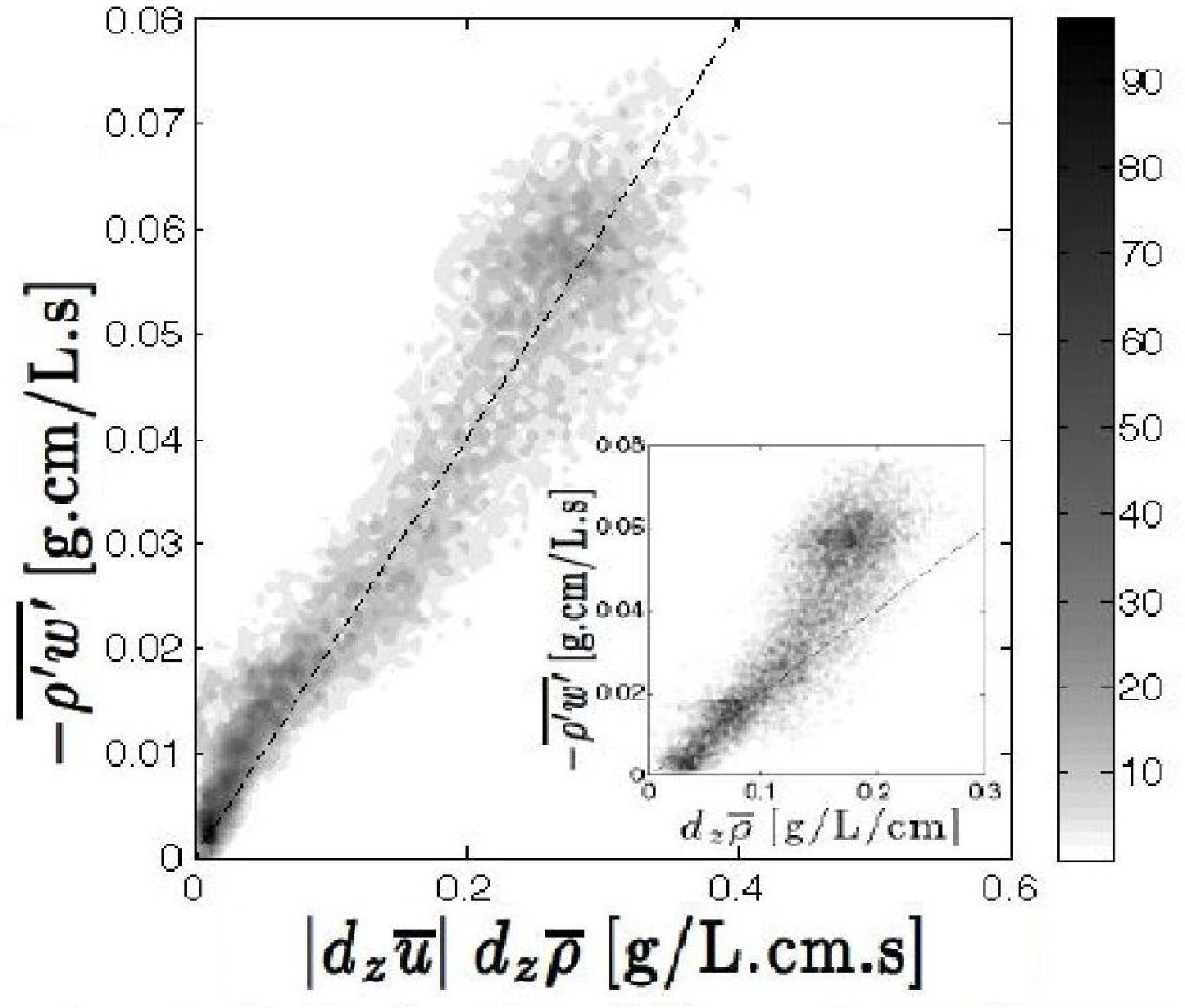}\\
\caption{\label{fig:Prandtl} Two dimensional histograms representing the correlation between turbulent stresses and mean gradients. Grayscale represents a number of entries in the histogram. (a) Momentum flux $\ol{u'w'}$ vs square of vertical velocity gradient $\dd_z \ol u^2$; (b) density flux $\ol{\rho'w'}$ vs product of vertical velocity and density gradient $\dd_z \ol u\dd_z \ol \rho$. The insets show a test of the linear assumption : $\ol{u'w'}$ vs $\dd_z \ol u$ in (a) and $\ol{\rho'w'}$ vs $\dd_z \ol \rho$ in (b). The respective slopes indicated by dashed lines yield respectively $L_m^2$ and $L_\rho^2$ for the main plots and $\nu_T$ and $\gamma_T$ for the insets.}
\end{minipage}}
\end{figure}

The simplest models used in geophysical models of climate (see, e.g., \cite{Chang}), assume a constant effective diffusivity, which yields a linear relation between the fluxes and the corresponding gradients. Because we are able to directly
measure $\ol{u'w'}$ and $\ol{\rho'w'}$, we can test the eddy diffusivity assumption.  In Fig.~\ref{fig:Prandtl} several two-dimensional histograms show the correlation between fluxes and gradients. Each entry in a histogram corresponds to one particular PIV grid point, averaged over time for a given experimental run and then over all experimental runs for the field view locations beyond $x=21$ cm. We exclude data from the vicinity of the solid boundary layer ($z<$1.5 cm) as well as data in the quiescent zone ($z>$7 cm), which display a somewhat different behavior.

In contrast to the constant eddy viscosity (diffusivity) assumption, the data in Fig.~\ref{fig:Prandtl}a show that the momentum flux is  proportional to the square of the velocity gradient. In the same way, Fig.~\ref{fig:Prandtl}b shows that the density flux is proportional to the product of the velocity gradient and the density gradient.  The quadratic dependence yields a nice description for the fluxes over most of the mean gradient range. The insets in Figs.~\ref{fig:Prandtl}a,b show the momentum and density fluxes versus the corresponding gradients. The data fails to support the constant eddy diffusivity model because the fluxes increase faster than the gradients. This qualitative trend was reported earlier \cite{Strang00}.

The quadratic behavior can be understood if both eddy diffusivities are linear functions of the velocity gradient, instead of being constant. Indeed, as the gradient becomes larger, one expects the turbulence intensity to increase. This can be interpreted in terms of a {\it mixing length} model, originally proposed by Ludwig Prandtl~\cite{Prandtl}. Prandtl's argument is analogous to that applied in the kinetic theory of gases to molecular transport processes: it assumes that the coefficient of eddy viscosity is equal to the product of a ``mixing length" $L_m$, characteristic of the mixing phenomena, and a suitable velocity: $\nu_T  \simeq  L_m\times U({\rm typical})$. One straightforward way of defining this velocity is to relate it to the mixing length and the mean velocity gradient along the direction of the transport: $U({\rm typical})=L_m\vert\dd_z \ol u\vert$. This relation assumes that $L_m$ is small enough so that the variation of the gradient over a distance $L_m$ can be neglected. The following expression for the momentum flux is then obtained: $\ol{u'w'}=-L_m^2\vert\partial_z \ol u\vert\partial_z \ol u$. In our case, where $\ol{u'w'}$ is positive and $\partial_z \ol u$ negative, this expression becomes: $\ol{u'w'}=L_m^2\partial_z \ol u^2$. The same argument for the density flux yields~(\cite{Hinze75}): $\ol{\rho'w'}=-L_{\rho}^2\vert\partial_z \ol u\vert\partial_z \rho$,
where $L_{\rho}$ is a mixing length associated with the density transport. As a result, averaging quantities along the downstream direction ($\langle\bullet\rangle_x$ symbol), we computed the mixing lengths as:

$$
L_m^2  =  \frac{\langle \ol{u'w'}\rangle_{x}}{\langle \dd_z\ol u^2\rangle_{x}}\;\;\; {\rm and~}
L_{\rho}^2  =  \frac{-\langle \ol{\rho'w'}\rangle_{x}}{\langle \vert\dd_z\ol u\vert\dd_z\ol\rho\rangle_{x}}$$

\begin{figure}
\includegraphics[width=6.5cm]{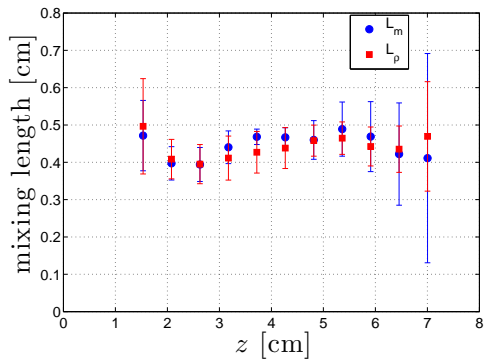}
\caption{\label{fig:lengths} (Color Online) Mixing lengths $L_m$ ($\bullet$) and $L_\rho$ ($\square$) vs distance $z$ from the plane. The error bars correspond to the standard deviation computed from the average along $x$.}
\end{figure}

The resulting $z$ profiles of mixing lengths are shown in Fig.~\ref{fig:lengths}. The profiles are fairly uniform, yielding mean values of $L_m=L_{\rho}=0.45\pm 0.1$ cm. The constancy of these lengths through the mixing zone and their equal magnitude (turbulent Prandtl number $\nu_T/\gamma_T\approx$ 1) yields a simple but powerful model for the mixing in stratified shear flows.  The physics of possible vertical variations of $L_m$ and $L_\rho$  is not discussed here.

By comparing the results presented above with data obtained with different conditions of stratification and/or turbulence level (although always at low $Ri$), we can draw some conclusions and provide a physical interpretation of the mixing length.  
Without turbulent enhancement, the mean $Ri$ is unchanged but the smaller $L_m\approx L_\rho \approx 0.3$ cm implies that
the degree of initial turbulence in the current strongly affects vertical mixing. Increasing the stratification leads as expected to less efficient mixing and a smaller mixing length ($L_m\approx L_\rho\approx$ 0.3 cm for a doubled density difference). Other data at higher Re suggest a stronger quadratic correlation between flux and gradient, as well as an increased mixing length ($L_m\approx$ 0.6 cm for $Re_\lambda=140$) but further experiments are necessary to confirm that trend. Finally, without stratification, i.e., for a free shear flow, the distribution of fluxes for a particular mean gradient is much wider, and neither a constant eddy viscosity model nor the mixing length model provides an adequate description of the data.

An interpretation of $L_m$ implies comparison with scales ~\cite{Smyth00} involving the competition of turbulent kinetic energy with the stabilization of buoyancy or the destabilization from shear.  The energy of a typical eddy of size $\ell$ is of order $v_t^2 \sim (\epsilon \ell)^{2/3}$ whereas the energy associated with buoyancy and shear is $v_b^2 \sim (N\ell)^2$ and $v_s^2 \sim (S\ell)^2$, respectively, where $N^2 = g\partial_z\ol\rho/\ol\rho$ and $S = \partial_z \ol u$.  Balancing the turbulent and forcing components yields a buoyancy length $L_o = (\epsilon/N^3)^{1/2}$ and a shear length $L_s = (\epsilon/S^3)^{1/2}$.  The smaller of these lengths limits the typical eddy size. We have $L_o \approx 2$ cm and $L_s \approx 0.5$ cm so we would expect $L_m \approx L_s$ as indeed we observe.  Further, the results obtained with different stratification and/or turbulence levels indicate $L_m \approx L_\rho \approx L_s$ for those conditions, implying that for low Richardson number the turbulent mixing lengths scale with shear rather than buoyancy.  Applying this argument to oceanic Mediterranean Overflow data~\cite{Price93} where one has measured values of $L_s~\approx$ 2.3 m and $\langle {\partial}_z u\rangle~\approx$ 0.013 ${\rm s}^{-1}$, our model predicts eddy diffusivities of $\nu_T\approx\gamma_T\approx L_s \langle {\partial}_z \ol u\rangle\approx$ 650 cm$^2$/s. We also expect that at higher Richardson number, as in the ocean, the length scale will be determined by buoyancy since $L_o < L_s$ so that our oceanic estimate may be a bit high.  Unfortunately, data allowing a direct comparison between our measurements and oceanic conditions is not available to our knowledge.  Parametrizations in ocean models~\cite{Lane_Serff00} have used values in the range 300$<\nu_T<$7000 cm$^2$/s for typical overflow scenarios. Critical to extrapolating to oceanic conditions is a systematic exploration of the dependence of the mixing lengths on turbulence intensity and on the degree of stratification as measured by $Ri$.

We acknowledge useful discussions with  P. Marcus, B. Wingate, M. Hecht, M. Lance, J.-F. Pinton, and C. Doering.  Work performed at Los Alamos National Laboratory was funded by the US Department of Energy under Contract No. DE-AC52-06NA25396.\\

\end{document}